\documentstyle[twocolumn,prl,aps,epsfig,graphicx]{revtex}

\begin{document}
\input{epsf}

\draft
\twocolumn[\hsize\textwidth\columnwidth\hsize\csname @twocolumnfalse\endcsname
\title{Order in the Heisenberg Pyrochlore: The
Magnetic Structure of Gd$_2$Ti$_2$O$_7$.}
\author{J.~D.~M.~Champion$^1$, A.~S.~Wills$^2$, T.~Fennell$^3$,
S.~T.~Bramwell$^1$,\\
J.~S.~Gardner$^4$, M.~A.~Green$^3$}

\address{$^1$ Department of Chemistry, University College London, 20
Gordon Street, London, WC1H~0AJ, United Kingdom.}
\address{$^2$ Institut Laue-Langevin, 6 Rue Jules Harowitz, BP 156, 38042
Grenoble Cedex 9, France.}
\address{$^3$ The Royal Institution of Great Britain, 21 Albemarle Street,
London, W1X 4BS, United Kingdom.}
\address{$^4$ Neutron Program for Materials Research, National Research
Council of
Canada, Chalk River, Ontario, Canada KOJ 1J0.}

\date{\today}
\maketitle

\begin{abstract}

The rare earth pyrochlore material Gd$_2$Ti$_2$O$_7$ is considered to be an
ideal model
frustrated Heisenberg antiferromagnet with additional dipolar interactions.
For this system
there are several untested theoretical predictions of the ground state
ordering pattern.
Here we establish the magnetic structure of isotopically enriched
$^{160}$Gd$_2$Ti$_2$O$_7$,
using powder neutron diffraction at a temperature of $50$~mK.  The magnetic
structure at
this temperature is a partially ordered, non-collinear antiferromagnetic
structure, with propagation
vector ${\bf k} = \frac{1}{2} \frac{1}{2} \frac{1}{2}$.
It can be described as a set of ``q=0'' ordered kagom\'e planes
separated by zero interstitial moments. This magnetic structure
agrees with theory only in part, leaving an interesting problem
for future research.

\end{abstract}
\vskip 0.2 truein
\newpage
\vskip2pc]

Lattices based on triangular motifs with antiferromagnetic nearest-neighbor
exchange are, in the absence of
further terms in the Hamiltonian, expected to be geometrically frustrated
as all the interactions between
coupled spins cannot be simultaneously minimized. This situation is at an
extreme in magnets based on the pyrochlore lattice, a face centred cubic
array of corner linked tetrahedra. Some thirty years ago Villain considered
the problem of the Heisenberg antiferromagnet on the pyrochlore lattice and
argued that it would be a ``cooperative paramagnet'' that remained
disordered and fluctuating down to zero temperature~\cite{Villain}. This
has subsequently been confirmed by both numerical and analytical
calculations~\cite{Reimers,RBS,Moessner,Canals}.  However, one question has
remained: do the dipolar interactions that must be present in a real
pyrochlore material play a decisive role in its low temperature physics? In
order to answer this question, interest has turned recently to the material
Gd$_2$Ti$_2$O$_7$, perhaps the only known pyrochlore material that is
sufficiently free of anisotropy and quenched disorder to realistically
approximate a dipolar Heisenberg system.

Confounding expectations, Gd$_2$Ti$_2$O$_7$ was found to have an
ordering transition at $0.97$ K, a temperature that is comparable
to the Curie-Weiss temperature, $9.6$~K~\cite{Raju}. To
understand this transition, two recent theoretical studies are
particularly important, both of which developed the methodology
of Reimers \textit{et al.}~\cite{RBS}. The first, by Raju \textit{et al.}~\cite{Raju},
considered a Heisenberg antiferromagnet with dipolar coupling and
showed that this does not completely lift the ground state
degeneracy to second order in the expansion of the free energy,
but rather stabilizes a degenerate set of periodic states with
$h~h~h$ propagation vectors. It was therefore suggested that the
transition observed in Gd$_2$Ti$_2$O$_7$ might be provoked by
thermal or quantum fluctuations, or alternatively by extra energy
terms in the spin Hamiltonian. The second study, by Palmer and
Chalker~\cite{Palmer}, in contrast, showed that the fourth order
term in the free energy expansion would select, from the
degenerate $h~h~h$ set, an ordered state with propagation vector
${\bf k} = 0$. In order to test these predictions we have now
determined the ordered magnetic structure of Gd$_2$Ti$_2$O$_7$ by
powder neutron diffraction techniques.

Natural Gd is almost opaque to neutrons due to the high
absorption cross section of $^{155}$Gd and $^{157}$Gd, so for
neutron studies one must use the isotope $^{160}$Gd.
Isotopically enriched $^{160}$Gd$_2$Ti$_2$O$_7$ was prepared from
$^{160}$Gd$_2$O$_3$ and TiO$_2$. Stoichiometric quantities were
mixed, ground, pressed into a pellet and fired at $1200-1400$~K
for $\sim 100$ hours. The grinding, pressing and firing processes
were repeated several times.  Powder neutron diffraction data were collected
using the POLARIS high intensity diffractometer at the ISIS
pulsed neutron facility. The sample was contained in a vanadium
can with $^3$He exchange gas and mounted in an Oxford Instruments
dilution refrigerator. Data were collected for $17$ hours above
and below the ordering transition (T$_c\sim 1$~K) at $5$~K, $500$~
mK and $50$~mK. A refinement of the pattern at
$5$~K showed the sample to be phase pure. Below T$_c\sim 1$~K new reflections were observed
that could be indexed with a magnetic propagation vector of
${\bf k} = \frac{1}{2} \frac{1}{2} \frac{1}{2}$.  The continuous
evolution of these reflections as a function of temperature
showed the transition to be second order. Given this observation,
representational analysis can be used to facilitate magnetic structure 
determination~\cite{Bertaut,Izyumov_book,Wills_Sarah,Wills_PRB_Jarosites}.

We define $G_{\mathbf k}$ as the little group of symmetry elements that
leave the propagation vector $\mathbf{k}$ invariant.
The magnetic representation, $\Gamma_{Gd}$, of the Gd site ($16$c in Fd$\bar3$m), can be
decomposed in terms of the irreducible representations (IRs) of $G_{\mathbf k}$.
These are listed in Table~\ref{basis_vector_table}
along with their associated basis vectors. The application of
$G_{\mathbf k}$ to the four Gd
positions of the tetrahedron of the asymmetric unit results in two orbits. When
applied to $0,0,0$ it
generates only the position $0,0,0$
--- orbit 1. When applied to another seed position, say
$\frac{3}{4},\frac{1}{4},\frac{1}{2}$~, it generates the remaining
$16$c sites:
$\frac{3}{4},\frac{1}{4},\frac{1}{2}$~
$\frac{1}{4},\frac{1}{2},\frac{3}{4}$~ and
$\frac{1}{2},\frac{3}{4},\frac{1}{4}$ --- orbit 2.

These two orbits can be understood in terms of a description of the
pyrochlore lattice as a set of
two dimensional kagom\'e lattice sheets (the $(1~1~1)$ planes) decorated by
interstitial spins which serve to link the sheets.
These form `up' and `down' pointing corner-sharing tetrahedra.
If the lattice is described in this way, then orbit 2 describes the three
atoms
that make up the triangular motif of a particular kagom\'e sheet and  orbit
1 is the unique atom present in the interstitial site.
The decomposition of the magnetic representation $\Gamma_{Gd}$ for orbits 1
and 2 are:
\begin{eqnarray}\label{decomp}\nonumber
&\Gamma_{Gd}^{orbit1} &= 0\Gamma_{1}^{(1)}+0\Gamma_{2}^{(1)}+1\Gamma_{3}^{(1)}+
0\Gamma_{4}^{(1)}+1\Gamma_{5}^{(2)}+0\Gamma_{6}^{(2)}
\\
&\Gamma_{Gd}^{orbit2} &= 0\Gamma_{1}^{(1)}+2\Gamma_{2}^{(1)}+0\Gamma_{3}^{(1)}+
1\Gamma_{4}^{(1)}+0\Gamma_{5}^{(2)}+3\Gamma_{6}^{(2)}.
\end{eqnarray}
Landau theory states that in simple systems, a continuous magnetic ordering
transition should involve only one IR becoming critical. However, in this case
the same IRs are not present in the decomposition of $\Gamma$ on the
different sites. For such a situation the only restriction that symmetry
affords is the constraint that each site, or orbit, orders under a single
IR with its own set of basis vectors. From the decompositions
(\ref{decomp}), there are two magnetic
representations for orbit 1 and three representations for orbit
2 (see Table~\ref{basis_vector_table}). Permutation of these results in six
allowed combinations of basis vectors, to be tested against the
experimental data.

Magnetic structure factors were calculated using the GENLES
routine of the GSAS suite~\cite{GSAS}, while the orientations of
the magnetic moments were controlled and refined separately by
the simulated annealing-based program SARA{\it h}--Refine for
GSAS~\cite{Wills_Sarah}.
 The $\chi^2$ and $R_{wp}$ 
\widetext
\begin{table}
\caption{Non-zero IRs and associated basis vectors $\psi_\nu$ for the space group Fd$\bar3$m with
${\bf k}= \frac{1}{2}\frac{1}{2}\frac{1}{2}$ calculated using the program SARA{\it h}-Representational Analysis.~\protect\cite{Wills_Sarah,Wills_PRB_Jarosites}. The labelling of the propagation vectors and the IRs follows the scheme used by Kovalev in his tabulated works~\protect\cite{Kovalev}.}
\begin{tabular}{cccccccccccccccc}
Orbit 1:& & & & & Orbit 2:\\    
  IR  &  BV   &  &  Atom 1 &  & IR& BV    & &   Atom 2 & & & Atom 3 &
  & &Atom 4 & \\
      &                &$m_x$ & $m_y$ & $m_z$& & &$m_x$ & $m_y$ &
      $m_z$&$m_x$ & $m_y$ &$m_z$  &$m_x$&$m_y$&$m_z$\\
\tableline
$\Gamma_3$& $\psi_1$&1&1&1& $\Gamma_2$   & $\psi_4$  &1 &1 &0 &0 &1 &1 &1 &0 &1\\
$\Gamma_5$&$\psi_2$ &$\bar{1}$&$\bar{1}$&0& & $\psi_5$  &0 &0 &1&1 &0 &0 &0 &1 &0\\
 &$\psi_3$&1&1&$\bar{2}$ & $\Gamma_4$   & $\psi_6$  &1 &$\bar{1}$ &0 &0 &1 &$\bar{1}$ &$\bar{1}$ &0 &1\\
& & & & &    $\Gamma_6$   & $\psi_7$  &1 &1 &0 &0 &$\bar{1}$ &$\bar{1}$ &$\bar{1}$ &0 &$\bar{1}$\\
& & & & &    & $\psi_8$  &0 &0 &1&$\bar{1}$ &0 &0 &0 &$\bar{1}$ &0\\
   & & & & & & $\psi_9$  &0 &0 & 0&0 &$\bar{1}$ &1 &$\bar{1}$ &0 &1\\
\end{tabular}
\label{basis_vector_table}
\end{table}
\narrowtext
\begin{figure}
\centerline{\hbox{\epsfig{figure=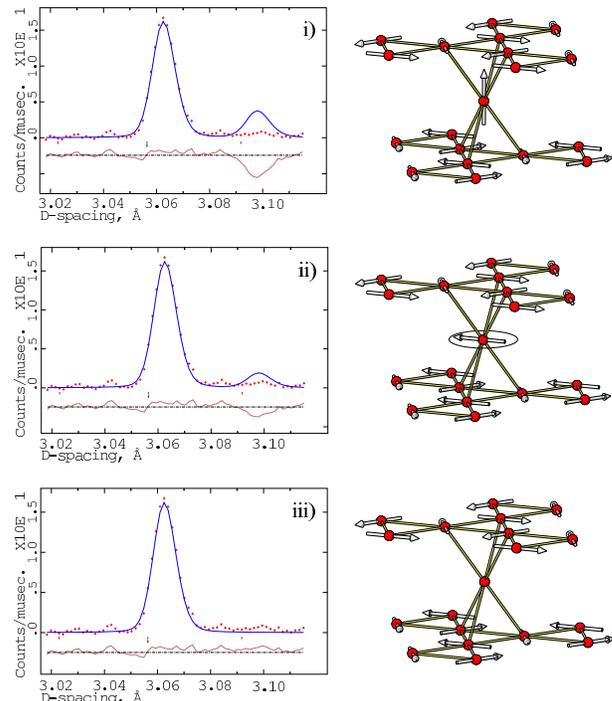,width=8cm}}}
\caption{The intensity of the magnetic reflection as a function of the
different structural
models (i), (ii), (iii) defined in the text, for the spin at position
$(0,0,0)$.}
\label{f_1}
\end{figure}

\noindent values for each refinement  are shown in Table~\ref{chitable}.
 It is clear that, among the six possibilities, there are only two candidates for the magnetic structure; all the other combinations can be immediately discarded.
 We refer to these candidate structures as (i) and (ii), see Table~\ref{chitable}. While both involve the basis vector $\psi_6$ for orbit 2, they differ in that, for
orbit 1, structure (i) involves $\psi_1$ while structure (ii)
involves the two basis vectors $\psi_2$ and $\psi_3$. The basis
vector $\psi_6$, which refers to the kagom\'e planes, in fact
corresponds to the ``q=0'' spin structure observed in ordered
kagom\'e antiferromagnets~\cite{q0kag}. For the interstitial
site, structure (i), with the basis vector $\psi_1$, corresponds
to a spin pointing parallel to the $[1~1~1]$ direction which is
perpendicular to the kagom\'e plane. Structure (ii), with basis
vectors $\psi_2$ and $\psi_3$, correspond to a spin lying with
any orientation in a plane parallel to the kagom\'e
\newpage

\noindent sheet. We
note that powder neutron diffraction cannot distinguish this
orientation. Three dimensional visualizations of the two possible
structures are shown in Fig.~\ref{f_1}. In both cases the
interstitial spins are ferromagnetically aligned within their
planes, and alternating planes are antiferromagnetically aligned.

Both structures (i) and (ii) fit well to the data, but there is an
important difference
at $\approx 3.1$ \AA (see Fig.~\ref{f_1}). Here, structure (i) predicts
more intensity than structure (ii), while
experiment suggests that the reflections at this d-spacing (for example
$\frac{3}{2}, \frac{3}{2}, \frac{5}{2}$)
are systematically absent. Thus neither structures (i) nor (ii) are
satisfactory.  A further fit was made which assigned zero moment to the
interstitial site, structure (iii). This is plausible, since a planar
antiferromagnetic structure on the kagom\'e sheets gives a zero mean field
at the interstitial site.

\begin{figure}
\begin{center}
\centerline{\hbox{\epsfig{figure=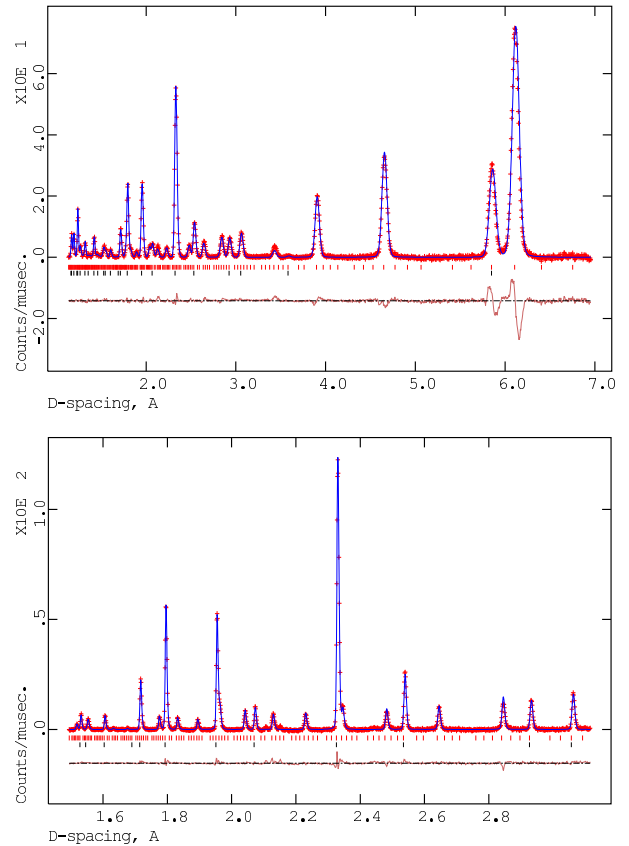,width=8cm}}}
\end{center}
\caption{The final refined magnetic profiles for data collected from the A and C detector banks of the POLARIS instrument from Gd$_2$Ti$_2$O$_7$ at 50~mK (nuclear reflections are marked by the lower line of ticks).}
\label{f_2}
\end{figure}

Structure (iii), with zero interstitial moment was found to
improve the fit to experiment by, most importantly, removing the
peak at $3.1$ {\AA} (see Fig.~\ref{f_1}). The final refined profile is shown in Fig.~\ref{f_2} and goodness of fit parameters are listed in Table~\ref{chitable}. We therefore conclude that structure (iii), with zero interstitial moment, is fully consistent with the experimental
data. It should be noted that the current diffraction experiment
cannot shed any light on whether the fourth moment is zero as a
result of static or fluctuating disorder.

In high symmetry crystals powder neutron diffraction does not
give an unambiguous determination of the magnetic structure,
because there is always the possibility of a multi-k structure.
In the present case, multi-k structures can be formed by
contributions from any of the four equivalent arms of the star of
${\bf k}$. However, recent investigations of Gd$_2$Ti$_2$O$_7$ by
M\"ossbauer spectroscopy~\cite{mossbauer} suggest that all the
magnetic spins lie in planes perpendicular to the $[1~1~1]$
direction below the ordering temperature. ESR measurements on the
system above $T_c$ have also indicated that the spins lie
preferentially in the $(1~1~1)$ plane~\cite{alia}. These
observations rule out a multi-k structure as the latter would
always have a finite spin component parallel to $[1~1~1]$ .
\begin{table}
\caption{Goodness of fit parameters $\chi^{2}$ and $R_{wp}$ for each
  of the six alternative combinations of IRs, for the data measured on
  the A and C detector banks of POLARIS. The last line indicates the
  fit that considers ordering only in orbit 2. The IRs corresponding
  to models (i), (ii) and (iii) are indicated.}
\begin{tabular}{ccccccc}
Label &\multicolumn{2}{c}{Orbit} &\multicolumn{2}{c}{A Bank} & \multicolumn{2}{c}{C Bank}  \\
 &1 & 2&$\chi^2$&$R_{wp}$&$\chi^2$&$R_{wp}$\\
\tableline
 &$\Gamma_3$&$\Gamma_2$&43.44 &0.0876&11.68&0.0424 \\
(i) &$\Gamma_3$&$\Gamma_4$& 3.844&0.0261&2.720&0.0205 \\
 &$\Gamma_3$&$\Gamma_6$& 13.17&0.0483&5.168&0.0282 \\
 &$\Gamma_5$&$\Gamma_2$& 40.76&0.0849&11.62&0.0423 \\
(ii) &$\Gamma_5$&$\Gamma_4$& 2.838&0.0224&2.494&0.0196 \\
 &$\Gamma_5$&$\Gamma_6$& 11.66&0.0454&4.957&0.0276 \\
\tableline
(iii)&     0     &$\Gamma_4$& 2.977&0.0229&2.186&0.0184 \\
\end{tabular}
\label{chitable}
\end{table}

Our result that ${\bf k} = \frac{1}{2} \frac{1}{2} \frac{1}{2}$
agrees with the prediction of Raju \textit{et al.} \cite{Raju}. These
authors considered a dipolar Heisenberg Hamiltonian with coupling
parameters appropriate to Gd$_2$Ti$_2$O$_7$. They showed that, to
quadratic order in the Landau expansion of the free energy, the
ground state consisted of a degenerate manifold of magnetic
structures with $h~h~h$ propagation vectors. It was therefore
suggested that thermal or quantum fluctuations, or additional
terms in the Hamiltonian, might operate to select a single
ordered state in the real material, with one particular value of
$h$. Palmer and Chalker extended this calculation to the fourth
order term, and found it to be minimized by setting moments of
equal magnitudes on each site of the elementary tetrahedron. A
general state was constructed by combining three basis vectors of the type
$\psi_6$ (in our notation). Each of these
basis vectors has three coplanar and one zero moment, as described
above, but differ in the choice of $\langle 1~1~1 \rangle$ plane.
The condition of equal moments was found to restrict the
possibilities to a single ground state with $h = 0$, with the four
spins in a $\langle 1~0~0 \rangle$ plane, contrary to our
findings.

There are two possible explanations for the disagreement between
experiment and theory. The first is that the real material has
additional terms in the spin Hamiltonian, such as further
neighbour exchange, that overcome the fourth order terms in the
free energy expansion. The influence of such terms was considered
in a rather general sense by Reimers \textit{et al.}~\cite{RBS} and it
would be interesting to develop the latter study in the present
context. From an experimental perspective, the Hamiltonian could
be investigated by studying relatives of Gd$_2$Ti$_2$O$_7$ such as
Gd$_2$Sn$_2$O$_7$ and Gd$_2$SbGaO$_7$, where additional energy
terms might have a different weighting. The second, and perhaps
more intriguing explanation, is that quantum fluctuations,
neglected in the classical theory, play a role in determining the
ordering pattern. However, the refined value of the ordered
moment of Gd$_2$Ti$_2$O$_7$ is 6.73$\,\pm\,0.05\,\mu_B$, which is
comparable with the maximum expected value of 7.94$\,\mu_B$. This
suggests that quantum mechanical fluctuations out of the ground
state are not very important in this spin $S = \frac{7}{2}$
system.

In conclusion, the model dipolar Heisenberg antiferromagnet
$^{160}$Gd$_2$Ti$_2$O$_7$ is partially ordered at $50$~mK in a pattern that
consists of
``q=0'' kagom\'e planes plus a zero interstitial moment. It would be
interesting to determine whether this moment exhibits a freezing transition
at some finite temperature. The partial order of chemically pure
Gd$_2$Ti$_2$O$_7$ contrasts with the behaviour of Heisenberg
antiferromagnets with small amounts of quenched chemical disorder, such as
CsNiCrF$_6$~\cite{Harris} and Y(Sc)Mn$_2$~\cite{YScMn2}. These materials show
spin freezing transitions in the absence of long range order and spin
correlations characteristic of the pure Heisenberg system~\cite{Canals,Harris,YScMn2}. Thus our results support the idea that quenched chemical
disorder is a strongly relevant perturbation in the pyrochlore system that
is responsible for the glassy ground states often observed~\cite{Peter}.
The theory of Raju \textit{et al.}~\cite{Raju} encompasses the experimental behaviour
of Gd$_2$Ti$_2$O$_7$ in that it predicts an $h~h~h$ propagation vector.
We note that, in symmetry terms, $h~h~h$ is an equivalence line in the
Brillouin zone, such that all possible ordering wavevectors share the IRs
and basis vectors that we have enumerated in Table~\ref{basis_vector_table}. The selection of $h =
\frac{1}{2}$ might be specific to this material or might be more general;
this question remains to be answered. The higher order theory of Palmer
and Chalker~\cite{Palmer} does not agree with our result.
It remains an interesting challenge, both experimental and theoretical
to understand the  origin of this disagreement.

\section*{acknowledgments}
The authors would like to thank M. J. Harris, R. Smith and R. Down for assistance with the
ISIS experiment, J. Hodges and J. P. Sanchez for communication and
discussion of
their M\"ossbauer results, A.~Hassan for communication and discussion of
the ESR result and C.~Lacroix, B.~Canals, P.~C.~W.~Holdsworth for
stimulating comments. JDMC and TF acknowledge funding by the EPSRC. ASW
would like to thank the Marie-Curie project of the EC for support.

\end{document}